\documentclass[aps,prb,reprint,noshowkeys,superscriptaddress]{revtex4-1}
\usepackage{graphicx,dcolumn,bm,xcolor,microtype,multirow,amscd,amsmath,amssymb,amsfonts,physics,mhchem,pifont,wrapfig,siunitx}

\usepackage[utf8]{inputenc}
\usepackage[T1]{fontenc}
\usepackage{txfonts}

\usepackage[normalem]{ulem}
\definecolor{darkgreen}{RGB}{0, 180, 0}

\usepackage[
	colorlinks=true,
    citecolor=blue,
    breaklinks=true
	]{hyperref}
\newcommand{\ie}{\textit{i.e.}}

\newcommand{\LCPQ}{Laboratoire de Chimie et Physique Quantiques (UMR 5626), Universit\'e de Toulouse, CNRS, UPS, France}
\newcommand{\VRIJE}{Department of Theoretical Chemistry and Amsterdam Center for Multiscale Modelling, FEW, Vrije Universiteit, De Boelelaan 1083, 1081HV Amsterdam, The Netherlands}

\newcommand{\cc}{\mu}

\begin{document}	

\title{Transient Uniform Electron Gases}

\author{Pierre-Fran\c{c}ois \surname{Loos}}
\email{loos@irsamc.ups-tlse.fr}
\affiliation{\LCPQ}
\author{Michael \surname{Seidl}}
\affiliation{\VRIJE}

\begin{abstract}
The uniform electron gas (UEG), a hypothetical system with finite homogenous electron density composed by an infinite number of electrons in a box of infinite volume, is the practical pillar of density-functional theory (DFT) and the foundation of the most acclaimed approximation of DFT, the local-density approximation (LDA). 
In the last thirty years, the knowledge of analytical parametrizations of the infinite UEG (IUEG) exchange-correlation energy has allowed researchers to perform a countless number of approximate electronic structure calculations for atoms, molecules, and solids.
Recently, it has been shown that the traditional concept of the IUEG is not the unique example of UEGs, and systems, in their lowest-energy state, consisting of electrons that are confined to the surface of a sphere provide a new family of UEGs with more customizable properties.
Here, we show that, some of the excited states associated with these systems can be classified as transient UEGs (TUEGs) as their electron density is only homogenous for very specific values of the radius of the sphere even though the electronic wave function is not rotationally invariant.
Concrete examples are provided in the case of two-electron systems.
\end{abstract}

\maketitle

\section{Uniform electron gases}
Alongside the two Hohenberg-Kohn theorems \cite{Hohenberg_1964} which put density-functional theory (DFT) on firm mathematical grounds and the Kohn-Sham (KS) formalism \cite{Kohn_1965} that makes DFT practically feasible, the uniform electron gas (UEG) \cite{Loos_2016} is one of the many pieces of the puzzle that have made DFT \cite{ParrBook} so successful in the past thirty years.
Indeed, apart from very few exceptions, most density-functional approximations are based, at some level at least, on the UEG via the so-called local-density approximation (LDA) \cite{Thomas_1927,Fermi_1927,Dirac_1930,Slater_1951,Ceperley_1980} which assumes that the electron density $\rho$ of an atom, a molecule, or a solid is locally uniform and has identical ``properties'' to the UEG with the same electron density.
A potential explanation of this success is that their on-top pair densities behave quite similarly. \cite{Perdew_1997}

Thanks to the construction of exchange-correlation LDA functionals \cite{Slater_1951,Vosko_1980,Perdew_1981,Perdew_1992,Chachiyo_2016} which can be loosely seen as a one-to-one mapping between a given value of the electron density and the exchange-correlation energy of the UEG, one can then straightforwardly compute, within KS-DFT, the electronic ground-state energy and properties of any molecules or materials with, nonetheless, a certain degree of approximation inherently associated with the approximate nature of the exchange-correlation LDA functional.
Moreover, one can also access excited states via the time-dependent version of DFT. \cite{Runge_1984,Casida_1995,Petersilka_1996,UllrichBook}

As commonly done, the LDA can be refined by adding up new ingredients, such as the gradient of the density $\nabla \rho$ [which defines the generalized gradient approximation (GGA)], \cite{Perdew_1986,Becke_1988,Lee_1988,Perdew_1996} the kinetic energy density $\tau$ (meta-GGA), \cite{Becke_1988b,Sun_2015} exact Hartree-Fock (HF) exchange (yielding the so-called hybrid functionals), \cite{Becke_1993a,Becke_1993b,Adamo_1999} and others. 
Each of these quantities defines a new rung of the well-known Jacob ladder of DFT \cite{Perdew_2001} that is supposed to bring electronic structure theory calculations from the evil Hartree world to the chemical accuracy heaven.

The UEG, also known as jellium in some context, \cite{Loos_2016} is a hypothetical infinite substance where an infinite number of electrons ``bathe'' in a (uniform) positively charged jelly of infinite volume. \cite{ParrBook,Loos_2016} 
It can be ``created'' via a \textit{gedanken} experiment by pouring electrons in an expandable box while keeping the ratio $\rho = N/V$ of the number of electrons $N$ and the volume of the box $V$ constant. 
In the so-called thermodynamic limit where both $N$ and $V$ goes to infinity but $\rho$ remains finite, the electron density eventually becomes homogeneous. In the following, this paradigm is named the infinite UEG (IUEG) for obvious reasons.
 
\section{Finite uniform electron gases}
Recently, it has been shown that one can create finite UEGs (FUEGs) by placing a finite number of electrons onto the surface of a sphere of radius $R$. \cite{Tempere_2002,Tempere_2007,Seidl_2007,Loos_2009a,Loos_2009c,Loos_2010e,Loos_2011b,Gill_2012,Loos_2018b} 
Of course, FUEGs only appear for well-defined electron numbers and electronic states. \cite{Rogers_2016,Rogers_2017}
In particular, the spin-unpolarized ground state of $N$ electrons on a sphere has a homogeneous density for $N = 2(\ell_\text{max}+1)^2$ (where $\ell_\text{max} \in \mathbb{N}$) for any $R$ values, and this holds also within the HF approximation. \cite{Loos_2011b}
This property comes from the addition theorem of the spherical harmonics \cite{NISTbook} $Y_{\ell m}(\bm{\Omega})$ (which are the spatial orbitals of the system in this particular case):
\begin{equation}
	\label{eq:ellMax}
	\sum_{\ell=0}^{\ell_{\rm max}} \sum_{m=-\ell}^{+\ell} Y_{\ell m}^*(\bm{\Omega}) Y_{\ell m}(\bm{\Omega}) = \frac{(\ell_\text{max}+1)^2}{2\pi^2},
\end{equation}
where $\bm{\Omega}=(\theta,\phi)$ gathers the polar and azimuthal angles, respectively.
Thanks to this key property, these FUEGs have been employed to construct alternative LDA functionals for both KS-DFT \cite{Loos_2014b,Loos_2017a} and ensemble DFT. \cite{Loos_2020g,Marut_2020}
Besides, hints of the equivalence of the FUEG and IUEG models have been found in the thermodynamic limit, \ie, when $\ell_\text{max} \to \infty$. \cite{Bowick_2002,Loos_2011b}

The case with $N=2$ electrons ($\ell_\text{max} = 0$) is of particular interest \cite{Seidl_2007,Loos_2009a,Loos_2018b} as it has been shown to be extremely useful for testing electronic structure methods \cite{Mitas_2006,Seidl_2007,Loos_2009a,Pedersen_2010,Loos_2012c,Schindlmayr_2013,Loos_2015b,Sun_2016,Loos_2018b} and is, furthermore, exactly solvable for a countably infinite set of $R$ values. \cite{Loos_2009c,Loos_2010e,Loos_2012}
In this case, the many-body hamiltonian reads
\begin{equation}
	\label{eq:H}
	\Hat{H} = \frac{\hat{\ell}_1^2 + \hat{\ell}_2^2}{2m\,R^2} + \lambda\frac{e^2}{r_{12}}.
\end{equation}
The squares $\hat{\ell}_i^2=-\hbar^2\hat{\Lambda}_i$ of the (orbital) angular momentum operators are essentially the angular parts $\hat{\Lambda}_i$ of the Laplacian,
\begin{equation}
\hat{\Lambda} = \frac{1}{\sin\theta}\pdv{}{\theta}\sin\theta \pdv{}{\theta} + \frac{1}{\sin^2\theta}\pdv[2]{}{\phi},
\end{equation}
while $r_{12}$ is the \textit{spatial distance} between the two electrons, \ie, the electrons interact Coulombically \textit{through} the sphere,
\begin{equation}
	r_{12} = R \sqrt{2(1 - \cos\gamma)} \equiv r_{12}(\gamma).
\end{equation}
Here, $\gamma=\gamma(\Omega_1,\Omega_2)$ is the angle between the two electrons on the sphere
(viewed from the spherical center),
\begin{equation}
	\label{eq:cosGamma}
	\cos\gamma = \cos \theta_1 \cos \theta_2 + \sin \theta_1 \sin \theta_2 \cos(\phi_1 - \phi_2).
\end{equation}
In Eq.~\eqref{eq:H}, we have introduced a coupling constant $\lambda$ which in the real universe (where the electrons, each with charge $-e$, \textit{repel} each other) has the value $\lambda=1$. However, in addition to this realistic situation, we here also wish to consider cases with $\lambda\ne1$, including the non-interacting case $\lambda=0$. In particular, we shall consider the interesting case $\lambda<0$ when the electrons \textit{attract} each other.

In atomic units ($m=e^2=\hbar=1$) where $R$ is given in units of the bohr radius $a_0=0.529$\AA, our hamiltonian reads
\begin{equation}
	\label{eq:H2}
	\Hat{H} = \frac1{R^2}\qty{ - \frac{\hat{\Lambda}_1+\hat{\Lambda}_2}2 + \cc \frac{1}{\sqrt{2(1 - \cos\gamma)}} },
\end{equation}
with an effective (dimensionless) coupling constant,
\begin{equation}
	\label{eq:cc}
\cc = \lambda R.
\end{equation}
Obviously, different interaction strengths $\lambda\ne0$ at a fixed radius $R>0$ are equivalent to different radii $R\ne0$ at 
a fixed interaction strength $\lambda>0$ (where a negative sign of $R$ has no geometric meaning but simply describes attractive 
electrons). 
Just as in the IUEG, the limit of high (low) density corresponds to the limit of weak (strong) interaction.

Following Breit, \cite{Breit_1930} one can write the total electronic wave function as
\begin{equation}
\label{eq:PhiOLD}
	\Phi(\bm{x}_1,\bm{x}_2) = \Xi(s_1,s_2) \, \chi(\Omega_1,\Omega_2) \, \Psi(\gamma),
\end{equation}
where $\Xi$, $\chi$ and $\Psi$ are the spin, the non-interacting angular and the interelectronic angular wave functions, respectively, and $\bm{x}_i = (s_i,\Omega_i)$ is a composite coordinate gathering the spin coordinate $s_i$ and the spatial (angular) coordinate $\Omega_i$ associated with the $i$th electron.

In the non-interacting limit $\lambda=\cc=0$, we have $\Psi(\gamma)=1$. In cases with finite interaction, $\cc\ne0$, this interelectronic wavefunction $\Psi(\gamma)=\Psi_{\cc}(\gamma)$, depending on the value of $\cc$, is either known analytically \cite{Loos_2009c,Loos_2010e,Loos_2012} or must be computed numerically.
  
The singlet and triplet spin wave functions read \cite{BetheBook}
\begin{subequations}
\begin{align}
	^1\Xi(s_1,s_2) & 
	= \frac{1}{\sqrt{2}} \, \Big[ \alpha(s_1) \beta(s_2) - \beta(s_1) \alpha(s_2) \Big]		\\
	^3\Xi(s_1,s_2) & = 
	\begin{cases}
		\alpha(s_1) \alpha(s_2),								\\
		\frac{1}{\sqrt{2}} \, \Big[ \alpha(s_1) \beta(s_2) + \beta(s_1) \alpha(s_2) \Big]	\\
		\beta(s_1) \beta(s_2).
	\end{cases}
\end{align}
\end{subequations}
Since the factor $\Psi(\gamma)$ in Eq.~\eqref{eq:PhiOLD} is symmetrical (upon swapping the coordinates of the two electrons), see Eq.~\eqref{eq:cosGamma}, the symmetry of the non-interacting angular wave functions $\chi(\Omega_1,\Omega_2)$ must be opposite to the one of the spin factor $\Xi(s_1,s_2)$. Consequently (as in the helium atom), the ground-state is a singlet,
\begin{equation}
	\label{eq:spin_1S}
	\chi_{^1S} (\Omega_1,\Omega_2) = Y_{00}(\Omega_1)Y_{00}(\Omega_2) = \frac1{4\pi},
\end{equation}
(in the usual notation $^{2S+1}L$ with the quantum number $L$ of total orbital angular momentum
$\hat{L} = \hat{\ell}_1+\hat{\ell}_2$ of the two electrons).
Similarly, for the $^3P$ and $^1P$ two-electron states, the angular non-interacting wave functions are \cite{Breit_1930,Seidl_2007,Loos_2009a,Loos_2009c,Loos_2010e}
\begin{subequations}
\begin{align}
	\label{eq:spin_3P}
	\chi_{^3P} (\Omega_1,\Omega_2) & = \frac1{\sqrt{2}}\,\Big[Y_{10}(\Omega_1)Y_{00}(\Omega_2)-Y_{00}(\Omega_1)Y_{10}(\Omega_2)\Big]
	\nonumber\\
    & = \frac{1}{4\pi}\sqrt{\frac32}\,\Big(\cos \theta_1 - \cos \theta_2\Big),
   	\\
	\label{eq:spin_1P}
	\chi_{^1P} (\Omega_1,\Omega_2) & = \frac1{\sqrt{2}}\,\Big[Y_{10}(\Omega_1)Y_{00}(\Omega_2)+Y_{00}(\Omega_1)Y_{10}(\Omega_2)\Big]
	\nonumber\\
    & = \frac{1}{4\pi}\sqrt{\frac32}\,\Big(\cos \theta_1 + \cos \theta_2\Big).
\end{align}
\end{subequations}

By definition, \cite{DavidsonBook} the total electronic density (as a function of the solid angle $\Omega$ on the sphere) is given by the integral
\begin{equation}
	\label{eq:rho}
	\rho(\Omega_1) = 2 \int \chi (\Omega_1,\Omega_2)^2 \Psi(\gamma)^2 d\Omega_2,
\end{equation}
(in the notation $d\Omega_2=\sin\theta_2\,d\theta_2\,d\phi_2$), where we have already integrated over the spin coordinates.

For the singlet ground state, we have $\chi_{^1S} (\Omega_1,\Omega_2) = \frac{1}{4\pi}$ [see Eq.~\eqref{eq:spin_1S}]. Consequently, 
\begin{equation}
\rho_{^1S}(\Omega_1) = \frac{2}{(4\pi)^2} \int \Psi(\gamma)^2 d\Omega_2 .
\end{equation}
Obviously, this integral cannot depend on $\Omega_1$, implying that the ground state of two electrons on the surface of a sphere has a uniform density,
\begin{equation}
\rho_{^1S}(\Omega) = \frac{2}{4\pi}, 
\end{equation}
for any value $\cc \in {\mathbb{R}}$ of the interaction constant. This result is also true for any excited states with $^1S$ symmetry.

For the other electronic states corresponding to higher total orbital angular momentum $L$, such as the lowest singlet and triplet $P$ states, \cite{Loos_2010e} the electron density is typically nonuniform, except in the very unlikely conditions discussed in the following section. \cite{Seidl_2007} 


\begin{figure}
	\includegraphics[width=\linewidth]{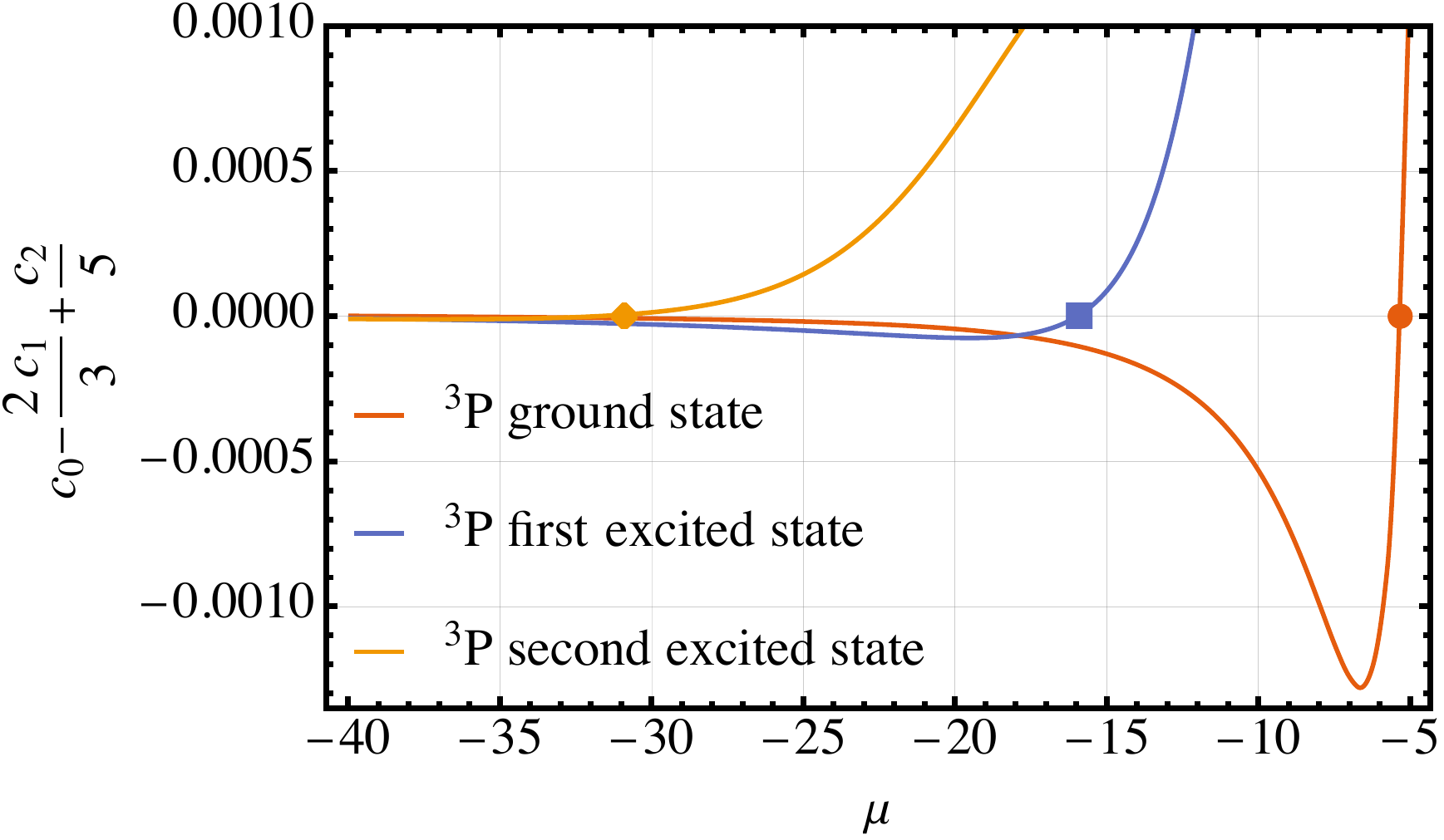}
	\caption{
	$c_0 - 2 c_1/3 + c_2/5$ as a function of $\mu$ for various states of $^3P$ symmetry.
	The zero associated with each state (which corresponds to the value of $\mu$ for which the electron density is uniform) is located by a marker.
	\label{fig:3P}}
\end{figure}

\section{Transient uniform electron gases}
As evidenced by Eq.~\eqref{eq:rho}, the electron density $\rho(\Omega)$ on the spherical surface is affected by both the non-interacting angular wave function $\chi(\Omega_1,\Omega_2)$ and the interelectronic one, $\Psi(\gamma)$. For particular values of $\cc$, a subtle interplay between these two quantities may result in a uniform density $\rho(\Omega)=\frac{2}{4\pi}$ as we shall illustrate now explicitly for the example of the $^3P$ and $^1P$ two-electron states.

To evaluate the integral of Eq.~\eqref{eq:rho} in the general case, we follow Ref.~\onlinecite{Seidl_2007} and decompose the square of the interelectronic wave function over the complete basis set composed by the Legendre polynomials $P_{\ell}(x)$,
\begin{equation}
	\label{eq:expansion_c}
	\Psi(\gamma)^2 = \sum_{\ell=0}^{\infty} c_{\ell} P_{\ell}(\cos \gamma).
\end{equation}
Using this expression along with the angular functions \eqref{eq:spin_3P} and \eqref{eq:spin_1P} for the $^3P$ and $^1P$ two-electron states (expressing their squares in terms of the $Y_{\ell m}(\Omega)$ and taking advantage of the general properties of spherical harmonics and Legendre polynomials), we find from Eq.~\eqref{eq:rho} the densities
\begin{subequations}
\begin{align}
	\label{eq:3P}
	\rho_{^3P}(\Omega) & = 
	\qty(c_0 - \frac{c_2}{5}) Y_{00}(\Omega)^2 
	+ \qty(c_0 - \frac{2 c_1}{3} + \frac{c_2}{5}) Y_{10}(\Omega)^2,
	\\
	\label{eq:1P}
	\rho_{^1P}(\Omega) & = 
	\qty(c_0 - \frac{c_2}{5}) Y_{00}(\Omega)^2 
	+ \qty(c_0 + \frac{2 c_1}{3} + \frac{c_2}{5}) Y_{10}(\Omega)^2.
\end{align}
\end{subequations}
Since $Y_{00}(\bm{\Omega})^2 = \frac1{4\pi}$, these densities are uniform if and only if the component associated with $Y_{10}(\bm{\Omega})^2$ vanishes,
\begin{equation}
	\label{eq:condition}
	c_0 \mp \frac{2 c_1}{3} + \frac{c_2}{5} = 0.
\end{equation}
From our numerical wave functions $\Psi(\gamma)$, the coefficients $c_{\ell}$ of the expansion \eqref{eq:expansion_c} can be extracted via
\begin{equation}
	c_{\ell} = \frac{2\ell+1}{2} \int_0^\pi P_{\ell}(\cos \gamma) \Psi(\gamma)^2 \sin \gamma d\gamma.
\end{equation}

For each (ground or excited) $^3P$ states, there exists one and only one value of $\cc$ in Eq.~\eqref{eq:cc}, $\cc=\cc_\text{UEG}<0$, for which these coefficients satisfy Eq.~\eqref{eq:condition}. $\cc_\text{UEG}$ can be computed numerically with great precision thanks to explicitly correlated calculations. \cite{Loos_2009a} Figure \ref{fig:3P} shows the behavior of $c_0 - 2 c_1/3 + c_2/5$ as a function of $\cc$ for the $^3P$ ground state and its first and second excited states. As one can see, $\cc_\text{UEG}$ is negative (which corresponds to an attractive ``electron'' pair \cite{Seidl_2007,Seidl_2010} or exciton \cite{Pedersen_2010,Loos_2012c}) and gets larger (in absolute value) for excited states.
Because this model has a nonuniform electron density except for a unique $\cc$ value, we name these ``ephemeral'' systems as transient UEGs (TUEGs). Note that this feature was first discovered in Appendix A of Ref.~\onlinecite{Seidl_2007}. There, also an estimate $\cc_\text{UEG} \approx -5.3$ was provided for the $^3P$ ground state, a rather good estimate that we refine here to $\cc_\text{UEG} \approx -5.32527$.

For the $^1P$ states, the condition $c_0 + 2 c_1/3 + c_2/5 = 0$ [see Eq.~\eqref{eq:condition}] cannot be fulfilled and, hence, these states never exhibit uniform densities. This further highlights the subtle balance that must be accomplished
between the non-interacting and the interelectronic (angular) parts (denoted here $\chi$ and $\Psi$, respectively) of the wave function
[see Eq.~\eqref{eq:rho}] and this can help us rationalizing why the $^3P$ states are TUEGs.

The $^3P$ non-interacting angular wave function, $\chi_{^3P}$, defined in Eq.~\eqref{eq:spin_3P} has the natural tendency to pull apart same-spin electrons in accordance with the Pauli exclusion principle, creating in the process a so-called Fermi hole. \cite{Boyd_1974,Giner_2016a}
The same physical effect can, independently, result from repulsion: In the case $\cc \gg 0$, two strongly repulsive electrons localize (or ``crystallize'') on opposite sides of the sphere to minimize their repulsion and they form a Wigner crystal. \cite{Wigner_1934} 
Oppositely, when $\cc \ll 0$, the two electrons are strongly attracted to each other, forming a tightly bound pair that moves freely on the sphere. \cite{Seidl_2007,Seidl_2010} For certain values $\cc<0$, the attractive force seems to exactly compensate the ``repulsive'' effect of the Pauli exclusion principle, thus making the total electron density uniform, hence producing TUEGs. In higher-energy excited states, the same-spin electrons are further away as compared to the ground state due to the larger number of nodes in the excited-state wave functions. Therefore, a compensating attraction must be larger, corresponding to larger negative values of $\cc$.

\begin{figure}
	\includegraphics[width=\linewidth]{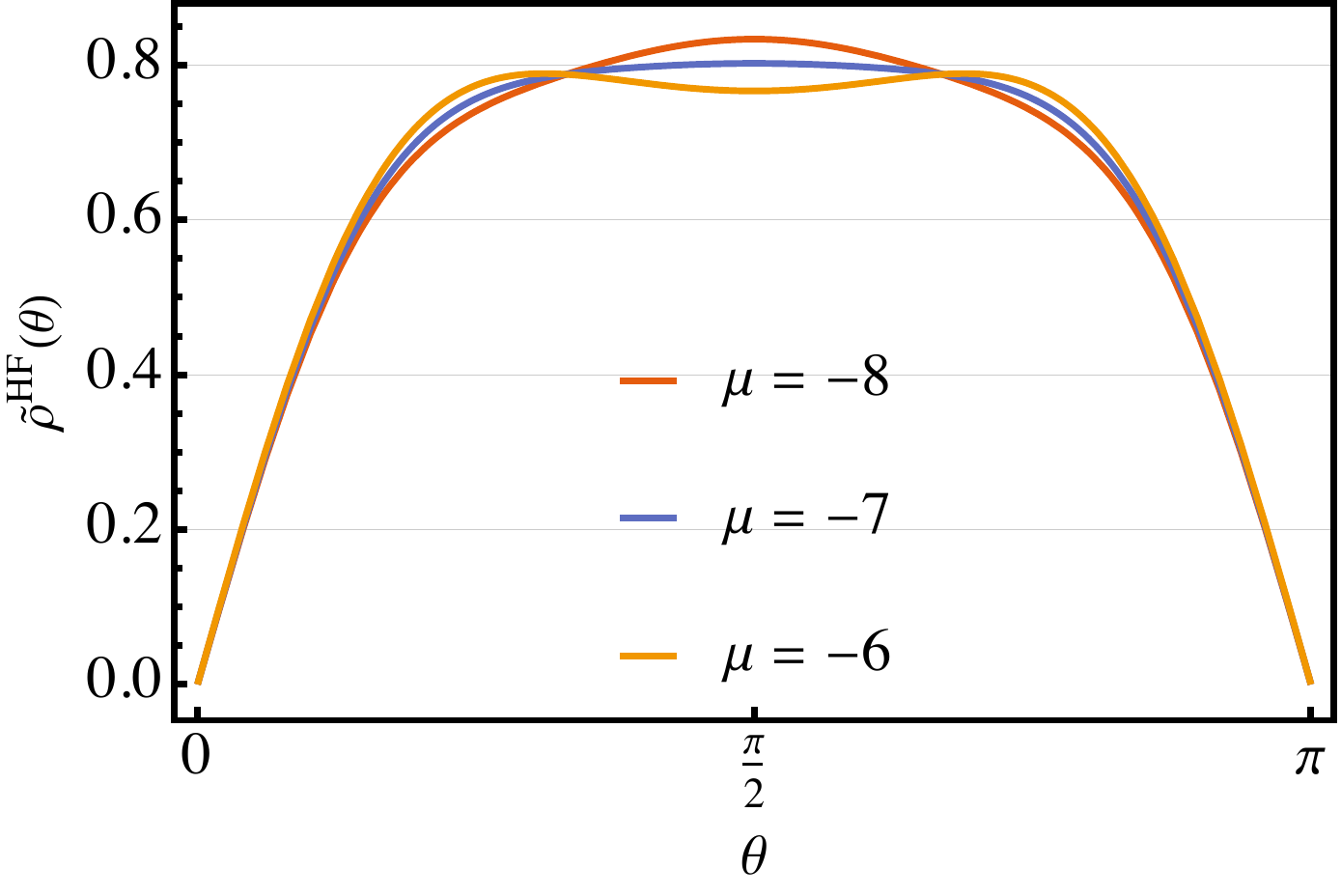}
	\caption{
	Hartree-Fock electron density $\tilde{\rho}^\text{HF}(\theta)$ as a function of the polar angle $\theta$ for various $\mu$ values.
	\label{fig:HF}}
\end{figure}

For the IUEG and FUEGs, the density is uniform independently of the level of theory, \ie, the system has homogeneous density within the exact theory or any approximate methods (such as the HF approximation) unless the spin and/or spatial symmetry is broken. \cite{Fukutome_1981,Stuber_2003,VignaleBook}
For TUEGs, however, the value of $\cc_\text{UEG}$ is, \textit{a priori}, highly dependent on the level of theory.
Indeed, it is very unlikely that the exact theory and the HF approximation provide the same value of $\cc_\text{UEG}$ as the uniformity stems from the competition between Fermi effects originating from the antisymmetric nature of the wave function (which are well described at the HF level) and correlation effects (which, by definition, are absent at the HF level). Actually, it is even possible for a system to be a TUEG within the exact treatment and being non-uniform for all $\cc$ values at the HF level. This seems to be the case for the present two-electron system.

Expanding the two HF orbitals of the $^3P$ ground state in a basis of zonal harmonics $Y_{\ell}(\theta) \equiv Y_{\ell 0}(\theta,\phi)$, we have not found any $\cc$ values for which the HF electron density,
\begin{equation}
\tilde{\rho}^\text{HF}(\theta) = \int_0^{2\pi}\rho^\text{HF}(\Omega) \sin\theta d\phi,
\end{equation}
is uniform.
At $\cc \approx -7$, however, $\tilde{\rho}^\text{HF}(\theta)$ is \textit{locally} uniform around $\theta = \frac{\pi}{2}$
(\ie, in a belt closely above and below the $xy$ plane), as shown in Fig.~\ref{fig:HF}. 
We believe that this outcome is a direct consequence of the single-determinant nature of the HF approximation which, by definition, can only include one (linear combination) of the three equivalent $sp$ configurations (\ie, $sp_x$,  $sp_y$, and $sp_z$). \cite{White_1971}
The fact that this phenomenon appears at larger (absolute) $\cc$ values in the HF approximation is not surprising as, contrary to the repulsive regime (\ie, $\cc > 0$) where the electrons are too close to each other at the HF level (compared to the exact picture), \cite{Gori-Giorgi_2008,Pearson_2009} in the attractive regime (\ie, $\cc < 0$) they are too far away from each other. 
This implies that the interaction strength has to be greater (which is equivalent to a larger absolute value of $\cc$) to overcome this drawback. \cite{Seidl_2010,Burton_2019a,Marie_2020}

\section{Concluding remarks}
Here, we have introduced the concept of transient UEGs (TUEGs), a novel family of electron gases that exhibit, in very particular conditions, homogenous densities.
Using the electrons-on-a-sphere model, we have presented an example of such TUEGs created thanks to the competing effects of the Pauli exclusion principle and the creation of an attractive electron pair. 
TUEGs with larger number of electrons certainly exist and we hope to investigate these in the future.
As a final remark, we would like to mention that a very similar analysis can be easily performed for higher-dimensional systems where TUEGs can likely be obtained for different $\mu$ values. \cite{Loos_2011b}
The three-dimensional version where electrons are confined to the surface of a 3-sphere (or glome) could be of particular interest, especially in the context of the development of new exchange-correlation functionals within DFT.\cite{Sun_2015,Agboola_2015,Loos_2017a}

\section*{Acknowledgements}
Stimulating discussions with Paola Gori-Giorgi are acknowledged.
This project has received funding from the European Research Council (ERC) under the European Union's Horizon 2020 research and innovation programme (Grant agreement No.~863481).
This work was also supported by the Netherlands Organisation for Scientific Research (NWO) under Vici grant 724.017.001.

%

\end{document}